\documentclass[letterpaper,12
pt]{article}
\usepackage{tabularx} 
\usepackage{amsmath} 
\usepackage{amsmath,amssymb}
\usepackage{amsthm}

\usepackage[bottom]{footmisc}
\usepackage{graphicx} 
\usepackage[margin=1in,letterpaper]{geometry} 
\usepackage{cite} 
\usepackage[final]{hyperref} 
\hypersetup{
	colorlinks=true,       
	linkcolor=blue,        
	citecolor=red,        
	filecolor=magenta,     
	urlcolor=blue         
}
\date{}
\DeclareUnicodeCharacter{2212}{-}
\usepackage{authblk}
\begin{document}
\title{\textbf{Geometric aspects of covariant Wick rotation}}
\author[1]{\Large{Raghvendra Singh} \footnote{raghvendra@imsc.res.in}}
\author[2]{\Large{Dawood Kothawala}  \footnote{dawood@iitm.ac.in}}
\affil[1]{\small{Institute of Mathematical Sciences, Homi Bhabha National Institute (HBNI), IV Cross Road, ~C. I. T. Campus, Taramani, Chennai 600 113, India}}
\affil[2]{\small{Department of Physics, Indian Institute of Technology Madras, Chennai 600 036}}
\maketitle
\begin{abstract}
\noindent We discuss the generic geometric properties of metrics $\widehat {g}_{ab}$ constructed from Lorentzian metric $g_{ab}$ and a nowhere vanishing, hypersurface orthogonal, timelike vector field $u^a$. The metric ${\widehat g}_{ab}$ has Euclidean signature in a certain domain, with the transition to Lorentzian signature occurring at some hypersurface $\Sigma$ orthogonal to $u^a$. Geometry associated with ${\widehat g}_{ab}$ has recently been shown to yield remarkable new insights for classical and quantum gravity. In this work, we prove several general results applicable in physically relevant spacetimes for congruences $u^i$ with non-zero acceleration $a^i$. We present as examples the cases of dynamical spherically symmetric spacetimes and spacetimes with maximal symmetry. We also investigate this formalism within the context of thermal effects in curved spacetimes with horizons. Specifically, we discuss: (i) the Holonomy of loops lying partially or wholly in the Euclidean regime. We show that the contribution of the Euclidean domain to holonomy is completely determined by extrinsic curvature $K_{ab}$ of $\Sigma$ and acceleration $a^i$. (ii) We also compute entropy using this formalism for simple field theories and obtain foliation dependent corrections for the Lanczos-Lovelock gravity, Bekenstein-Hawking entropy relation in four spacetime dimensions.
\end{abstract}

\section{Introduction}
The conventional method of Wick rotation, which involves the transformation $t \to  i t$ is known to be problematic when applied to the metric tensor itself since the procedure does not always produce {\it real} Euclidean metrics, and the interpretation of {\it imaginary} part of the metric is quite ambiguous. The flat spacetime provides us with a preferred choice of the time coordinate i.e. the one used by inertial observers but there is no such preferred choice available in a general curved spacetime. Moreover, the transformation $t \to i t$ is not covariant as it stands but for the interpretation of physical effects usually associated with Euclideanization, such as thermal properties of horizons and tunneling amplitudes, it is desirable to have manifest covariance. 
The above issues are best demonstrated in the case of non-stationary metrics, as well as stationary metrics with off-diagonal ``time-space'' components. Such oddities are easily illustrated with a simple example of de sitter metric in two different coordinate systems. In the {\it positive spatial curvature slicing}, the metric is 
\begin{align}\label{Ex1}
   ds^2=-d\tau^2+\cosh^2\tau d\Omega^2_3 \,,
\end{align}
continuation to imaginary time gives us spherical coordinate on $\textbf{S}^4$
\begin{align}
    ds^2=d\tau^2+\cos^2\tau d\Omega^2_3=d\Omega^2_4\,.
\end{align}
We may consider $\tau$ as angular coordinate with $ \tau \in [-\pi/2,\pi/2]$\cite{EQG}. On the other hand, in the \textit{negative spatial curvature slicing}, the metric is
\begin{align}\label{Ex2}
    ds^2=-d\tau^2+\sinh^2\tau d \mathbb{H}^2_3\,,
\end{align}
with $d \mathbb{H}^2_3$ the line-element on a unit hyperboloid. Analytic continuation yields 
\begin{align}
    ds^2=d\tau^2-\sin^2\tau d \mathbb{H}^2_3\,.
\end{align}
Which has signature (3,1)! It should be clear that conventional Wick rotation through imaginary time does not guarantee any unique structure for the corresponding geometry.

As mentioned above, much of the above oddities and ambiguities are tied to a lack of manifest covariance in the standard analytic continuation of the time coordinate. A covariant alternative to Wick rotation can indeed be given if one introduces an observer field $u^a$, which is essentially a non-vanishing timelike field associate with the original Lorentzian spacetime $(M, g_{ab})$. Let $\lambda$ be the parameter along $u^a$, and consider the class of metrics 
\begin{eqnarray}
    \widehat g^{a b}=g^{a b}-\Theta u^a u^b\,,
    \label{eq:ghat}
\end{eqnarray}
with an arbitrary function $\Theta$ that smoothly goes from $\Theta= −2$ to $\Theta = 0$, with signature of $\widehat g^{a b}$ going from Euclidean to Lorentzian respectively. We take $\widehat g^{a b}$ as the candidate metric that has a Euclidean regime for $\Theta < −1$ and Lorentzian regime for  $\Theta > −1$, while being degenerate for $\Theta = −1$. We call the co-dimension one hypersurface defined by $\Theta = −1$ as $\Sigma_0$. The above formalism was given in \cite{Dawood, Dawood2}, motivated essentially by an observation in Hawking and Ellis\cite{Ellis} (which corresponds to purely Euclidean metrics with $\Theta=-2$). It goes beyond the conventional constructions which aim to obtain Euclidean counterparts of Lorentzian geometries because it describes geometries with both Euclidean and Lorentzian regimes. Several new features arise in the above formalism which are not present in the conventional Wick rotation, including terms that have compact support on $\Sigma_0$. We refer the reader to \cite{Dawood} for a more detailed discussion relevant from the context of Euclidean quantum gravity and \cite{Dawood2} for a discussion on how it results in a Euclidean action with interesting mathematical structure.

We can immediately apply this construction to the examples (\ref{Ex1}) and (\ref{Ex2}) discussed above, which should already highlight the key features and differences from conventional case. For both of these cases, choose $u^a=(1,0,0,0)$ as the direction field. Then, we obtain following metrics for the positive and negative spatial slicing respectively 
\begin{eqnarray}
    ds^2 &=& -\frac{1}{1+\Theta}d\tau^2+\cosh^2\tau d\Omega^2_3 \,,
    \\
    ds^2 &=& -\frac{1}{1+\Theta}d\tau^2+\sinh^2\tau d\mathbb{H}^2_3 \,.
\end{eqnarray}
Unlike usual Wick rotation, we get here metrics with a well defined Euclidean regime (corresponding to $\Theta<-1$).

The two previous works mentioned above \cite{Dawood},\cite{Dawood2}  studied the geometric aspects of curvature associated with geodesic congruences (characterize freely falling frames) in well known spacetimes \cite{Dawood} and the implications for Euclidean action and quantum gravity \cite{Dawood2}. Given that Euclidean methods have most prominently been used in the study of thermal properties associated with the presence of horizons, in this paper we probe the above formalism from this point of view, focusing on features that arise for accelerated observer congruence, including cases when the congruence is not hypersurface orthogonal. In particular, we highlight the results for the case where $u^a$ is along a timelike Killing vector field of a given spacetime. We also exhibit the full structure of the Kretschmann scalar and the Weyl tensor, which should be useful in the physical interpretation of the Euclidean domain of $\widehat{g}_{ab}$. Motivated by a recent result by Samuel \cite{Samuel} based on similar consideration, we also analyse the interesting case of {\it holonomy} associated with loops that cross the hypersurface $\Sigma_0$, having one part in the Euclidean domain and rest in the Lorentzian one. Finally, we apply the formalism to compute the black hole entropy that leads us to the new and enthralling set of results. 

\section{The Curvature Tensors Associated With \texorpdfstring{$\widehat{g}$}{\265}}

It is a lengthy, though straightforward exercise to compute the various geometrical quantities associated with the metric $\widehat{g}_{ab}$ (\ref{eq:ghat}). Some of the basis quantities are given in Appendix \ref{app:data1}. Our focus here is to present the associated curvature tensor and its concomitants. This was done in an earlier work \cite{Dawood}, but under the assumption that the congruence is geodesic. We will here relax this assumption. In addition, we also give the expressions for the Kretschmann invariant and the Weyl tensor associated with $\widehat g$, since these are directly relevant from the point of view of applications.

Using the results from \ref{app:data1}, we obtain the curvature tensor associated with $\widehat{g}$ in the terms of the quantities associated with $g$ and those describing the intrinsic and extrinsic geometry of hypersurfaces foliated by $\textbf{u}$.

The Riemann tensor turns out to be 
\begin{align}\label{Riemann}
 \widehat R_{ab}^{\ \ c d} &=& 
R_{a b}^{\ \ c d} + 2\Theta \left(-u^{[c}R_{a b m}^{\ \ \ \ d]}u^m-K^{\ d}_{[a}K_{b]}^{\ c}+2t_{[a}a_{b]}a^{\textbf{[}c}u^{d\textbf{]}}+2u^{\textbf{[}c}(\nabla _{[a}a^{d\textbf{]}})t_{b]} \right) +2\dot \Theta u^{\textbf{[}c}K_{[a}^{\ d\textbf{]}}t_{b]}\,.  
\end{align}


We may similarly write down the expressions for Ricci and Einstein tensors and the Ricci scalar. We quote the final expressions below:
\begin{eqnarray}\label{Ricci}
\widehat R^a_{\ c} &=& (1+\Theta) R^a_{\ c}
- \Theta\left( ~^{(3)}R^a_{\ c} - t_c C^a + t_ca^b K^a_{\ b} - a^a a_c - g^{la}h^r_l\nabla_r a_c + u^a t_c\nabla_b a^b \right)
\nonumber \\
&& \hspace{2.05cm} + \; (1/2) {\dot\Theta} \left(  \pi^a_{\ c}+K\delta^a_{\ c}\right)\,,
\\
\widehat R&=&(1+\Theta)R+\Theta \left(-~^{(3)}R+2\nabla_b a^b\right)+\dot \Theta K\,,\label{Ricci2}
\\
\widehat G^a_{\ c} &=& 
(1+\Theta) G^a_{\ c} -
\Theta \left(~^{(3)}G^a_{\ c}+(1/2)~^{(3)}R  u^a t_c-t_c C^a-t_c a^b K^a_b-a^a a_c+u^a t_c\nabla_b a^b-g^{l a}h^r_l\nabla_r a_c\right)
\nonumber\\
&& \hspace{2.05cm} + \; (1/2) {\dot \Theta} \pi^a_{\ c} \,,
\end{eqnarray}
where we have used Gauss-Codazzi and Gauss-Weingarten equations, $C^m=D_a K^{a m}−D^m K$, with $D_m$ the natural covariant derivative that acts on tangent vectors to the hypersurfaces $\Sigma_t$, and $\pi^a_{\ b}=K^a_{\ b}-K h^a_{\ b}$, $h_{a b}$ being the induced metric on $\Sigma_t$.

Next we discuss some quantities of direct physical significance that can be immediately constructed from the above expressions. In particular, we quote the expressions for the tidal part of the Riemann tensor, Kretschmann scalar and the Weyl tensor associated with $\widehat g$. These expressions were not given in the closed form in previous literature but are expected to be of obvious significance from the point of view of physical interpretation of the geometry described by $\widehat g$.

\subsection{Tidal tensor}

From the above, we can immediately write down the components of the Tidal part of the Riemann tensor, defined by
$E^i_{\ d}:=R^i_{\ b c d}u^b u^c$ 
\begin{align}\label{tidal tensor}
    \widehat{E}^i_{\ d}=E^i_{\ d}+F\left(g^{a i}\nabla_a a_d +u^i\nabla_{\vec{u}}a_d-t_d a_c K^{i c}-a^i a_d\right)+\frac{\dot\Theta}{1+\Theta}K^i_{\ d}\,,
\end{align}
where $F={\Theta}/({1+\Theta})$ and $\nabla_{\vec{u}}a_d = u^k \nabla_{k} a_d$. Let us consider $\xi^i$ be a vector orthogonal to $t_i$ (dual of $u^i$ \ref{app:data1} ), so that $\xi^i t_i=0$. This vector could, for example, represent deviation between members of the congruence $u^i$. From the above expression for tidal tensor, it immediately follows that
\begin{align}\label{tidal tensor2}
    {\widehat {\cal A}}^i = \widehat{E}^i_{\ d} \; \xi^d = { {\cal A}}^i + 
    F \left(g^{a i} \xi^d \nabla_a a_d + \xi^d u^i\nabla_{\vec{u}} a_d - a^i \xi^d a_d\right)+\frac{\dot\Theta}{1+\Theta}K^i_{\ d} \; \xi^d\,,
\end{align}
where ${{\cal A}}^i = E^i_{\ d} \; \xi^d$. The component of ${\widehat{\cal A}}^i$ orthogonal to $u^i$ is then given by ${\widehat{\cal A}_{\perp}}^i = {\widehat{\cal A}}^i + ( {\widehat{\cal A}}^k t_k ) u^i$, and quickly checking that ${\widehat{\cal A}}^k t_k={{\cal A}}^k t_k$, we obtain
\begin{align}\label{tidal tensor3}
    {\widehat {\cal A}_\perp}^i = { {\cal A}_\perp}^i + 
    F \left(h^{a i} \xi^d \nabla_a a_d - a^i \xi^d a_d \right) + \frac{\dot\Theta}{1+\Theta}K^i_{\ d} \; \xi^d\,,
\end{align}
where $h^{ai} = g^{ai}+u^a u^i$ is the standard projector. The astute reader would have noticed that the quantity ${ {\cal A}_\perp}^i$ we have constructed above is precisely the deviation acceleration associated with the congruence when $a^i=0$. For an accelerated congruence, one needs to consider the Fermi acceleration, which can be easily done but we skip it. What is worth noticing here is that in the Euclidean regime $(\Theta=-2, F=2)$, for non-geodesic congruences, there is already an additional term in the deviation acceleration solely due to the signature change of the metric. Of course, to extract a direct physical measure of this acceleration, one must properly take into account the normalization of vectors in the Euclidean sector as well, but this is straightforward and we do not state it here.
\subsection{Kretschmann scalar}
Kretschmann scalar(let us denote it by S)  can be express in the following fashion.
\begin{align}\label{S}
  \widehat{S}=S &+\Theta\left(8R_{a b}^{\ \ c d}u^{[a}\nabla_{[c}K_{d]}^{\ b]} -4 R_{a b}^{\ \ c d}K_{[c}^{\ b}K_{d]}^{\ a}\right)\nonumber\\
 &+4\Theta^2\Large{\textbf{(}}(\nabla_{\vec{u}}K_b^{\ d})(\nabla_{\vec{u}}K^b_{\ d})+2(\nabla_{\vec{u}}K^b_{\ d})K^{a d}K_{b a}+K^{d b}K_{ c d}K^{a c}K_{b a}\nonumber\\
 &~~~~~~~~~~~+\frac{1}{2}(K_{m n}K^{m n})^2-\frac{1}{2}K^{c b}K_{a c}K^{d a}K_{b d}\Large{\textbf{)}}\nonumber\\
 &+2\Theta\dot\Theta\left(K^b_{\ c}\nabla_{\vec{u}}K_b^{\ c}+K^{a c}K_{b a}K^b_{\ c}\right) +2\dot\Theta u^{[a}K^{b]}_{\ [c}t_{d]}R_{a b}^{\ \ c d}+\dot{\Theta}^2K^{b d}K_{b d}\,.
\end{align}
\subsection{Weyl Tensor}
Writing  the expression for  Weyl tensor is much more tedious, though we write 4 dimensional Weyl tensor using the equations (\ref{Riemann}, \ref{Ricci}) as follows,
\begin{align}
\widehat{W}_{a b c d}=W_{a b c d}&+\Theta \huge{\textbf{(}}\frac{1}{3}(K^2-K_{m n}K^{m n}+R+2 \nabla_m a^m-2R_{m n}u^m u^n)(g_{a[c}g_{d]b}+2F g_{\textbf{[}a[c}t_{d]}t_{b\textbf{]}})\nonumber\\
&~~~~~~+\frac{2R}{3(1+\Theta)}g_{\textbf{[}a[c}t_{d]}t_{b\textbf{]}}-\frac{2}{1+\Theta}t_{\textbf{[}a} t_{[c}R_{d]b\textbf{]}}-2 Kg_{\textbf{[} a[c}K_{d]b\textbf{]}}-2g_{\textbf{[} a[c}u^m\nabla_m K_{d]b\textbf{]}}\nonumber\\
&~~~~~~+\frac{2}{1+\Theta}g_{\textbf{[} a[c}t_{d]}t_{b\textbf{]}} \nabla_l a^l-2a^m(K_{m\textbf{[}c}g_{d\textbf{]}[a}t_{b]}+K_{m [a}g_{b]\textbf{[}c}t_{d\textbf{]}})-2F Kt_{\textbf{[} a}t_{[c}K_{d]b\textbf{]}}\nonumber\\
&~~~~-2Ft_{\textbf{[} a}t_{[c}n^l\nabla_l K_{d]b\textbf{]}}-2K_{d[a}K_{b]c}
+\frac{8}{1+\Theta}t_{\textbf{[}c}(\nabla_{[a}a_{d\textbf{]}})t_{b]}+\frac{8}{1+\Theta} t_{\textbf{[}d} t_{[a}a_{b]}a_{c\textbf{]}}\huge{\textbf{)}}\nonumber\\
&-\dot \Theta \left(-\frac{K}{3}g_{a[c}g_{d]b}-\frac{2\Theta+3}{3(1+\Theta)}K g_{\textbf{[}a[c}t_{d]}t_{b\textbf{]}}+g_{\textbf{[}a[c}K_{d]b\textbf{]}}+\frac{2+\Theta}{1+\Theta}K_{\textbf{[}a[c}t_{d]}t_{b\textbf{]}}\right)\,,
\end{align}
where we have used anti symmetric index notation e.g. $K_{\textbf{[}a[c}t_{d]}t_{b\textbf{]}}=-\frac{1}{4}(K_{a c}t_d t_b -K_{a d}t_c t_b+K_{b d} t_c t_a-K_{b c} t_d t_a)$. The above expression for the Weyl tensor looks complicated, so we try to write it in simpler form by the following expression
\begin{align}\label{Weyl}
\widehat{W}_{a b}^{\ \ c d}=W_{a b}^{\ \ c d} &+\Theta \huge{\textbf{(}}4u^{[c}\nabla_{[a}K_{b]}^{\ d]}-2K_{[a}^{\ d}K_{b]}^{\ c}-2\delta_{[a}^{\ [c}u^{d]}\nabla_{b]}K
+2\delta_{[a}^{\ [c}u^{d]}\nabla_m K_{b]}^{\ m}\nonumber\\
 &~~~~~~~+2\delta_{[a}^{\ [c}u^m\nabla_{b]}K_m^{\ d]}-2\delta_{[a}^{\ [c}u^m \nabla_mK_{b]}^{\ d]}+2\delta_{[a}^{\ [c}K_{b]}^{\ m}K_m^{\ d]}-2\delta_{[a}^{\ [c}K_{b]}^{\ d]}K\nonumber\\
 &~~~~~~~+\frac{1}{3}\delta_{[a}^{\ c}\delta_{b]}^{\ d}(2 u^m\nabla_m K+K_{m n}K^{m n}+K^2)\huge{\textbf{)}}\nonumber\\
& +\dot\Theta \left(2u^{[c}K_{[a}^{\ d]}t_{b]}+\delta_{[a}^{\ [c}t_{b]}u^{d]}K-\delta_{[a}^{\ [c}K_{b]}^{\ d]}+\frac{1}{3}K\delta_{[a}^{\ c}\delta_{b]}^{\ d}\right)\,.
\end{align}

The above expression clearly shows that, in general, a conformally flat geometry $g$ will not be mapped to a conformally flat $\widehat g$, the additional terms being characterised by extrinsic curvature of the hypersurfaces orthogonal to $u^a$. It will be interesting to understand the consequences of this property, specifically in the context of early universe cosmology. From this point of view, let us consider the illustrative example of the standard FLRW geometry
\begin{align}
ds^2=-dt^2+a^2(t)d\Omega^2_{(k)}~~~(k=-1,0,1)\,,
\end{align}
where $a(t)$ is scale factor. We choose as our congruence the vector field is $t_m=-\partial_m t$. A quick calculation gives $K_m^{\ n}=\nabla_m u^n=({\dot a}/{a}) h_m^n, K={3\dot a}/{a}$ and plugging this in  equation (\ref{Weyl}) gives
\begin{align}
    \widehat{W}_{a b}^{\ \ c d}=W_{a b}^{\ \ c d}=0\,.
\end{align}
The above result would most easily be obtained by writing down $\widehat g$ and noticing that it is easily put in a conformally flat form. However, $\widehat{W}_{a b}^{\ \ c d}$ will be non-vanishing in the Euclidean regime of FLRW for an arbitrary $u^a$. As stated above, it will be interesting to extract physical significance of this in the context of quantum cosmology.

\subsection{Foliations with vanishing extrinsic curvature}

In physically relevant applications of Euclidean methods, foliations with vanishing extrinsic curvature play a particularly significant role. Under $t \to i t$ in conventional Wick rotation, since $K_{ab} \to i K_{ab}$, the matching of a Euclidean domain to Lorentzian one, is done on a surface of vanishing extrinsic curvature. The formalism presented here does not a priori require any constraint on $u^a$, therefore allows for $K_{ab}$ to be non-zero everywhere. Nevertheless, we will now show that the results match with conventional Wick rotation for a foliation by hypersurfaces with $K_{ab}=0$. This will also immediately apply to foliation by static timelike Killing vector fields whose extrinsic curvature vanish. 
\\
\\
\noindent \textbf{Claim:}\label{Claim-1} \textit{If one chooses a smooth vector field {$u^i$} in such a way that its level surfaces foliate the spacetime by non-intersecting extrinsically flat hypersurfaces, then following identities hold:}
\begin{align}\label{Identities}
 \widehat{R}_{a b}^{\ \ c d} = R_{a b}^{\ \ c d}\,,~~\widehat{R}^a_{\ c}={R^a_{\ c}}\,,~~\widehat R=R\,,\nonumber\\
   \widehat{G}^a_{\ c}=\widehat{G}^a_{\ c}\,,~~\widehat{S}=S\,,~~\widehat{W}_{a b}^{\ \ c d}={W_{a b}^{\ \ c d}}\,.
\end{align}
\textbf{Proof:}
For $K_{ab}=0$, we use Gauss-Codazzi and Gauss-Weingarten equations to write
$~^{(3)}R^a_{\ c}$ as
\begin{align}\label{3R}
    ~^{(3)}R^a_{\ c}=h^{a l}h^m_{\ l} h^b_{\ c}R_{m b}+h^{l a}h_l^{\ m}h_c^{\ b}R_{m n b d}u^n u^d\,.
\end{align}
We simplify the RHS of the above equation as
\begin{align}\label{3R1}
    h^{a l}h^m_{\ l} h^b_{\ c}R_{m b}=R^a_{\ c}-t_c u^a \nabla_m a^m\,,
\end{align}
\begin{align}\label{3R2}
    h^{l a}h_l^{\ m}h_c^{\ b}R_{m n b d}u^n u^d=a^a a_c +h^r_{\ j} g^{j a}\nabla_r a_c\,.
\end{align}
Substituting the equations (\ref{3R}-\ref{3R2}) into the equation (\ref{Ricci}), we get
\begin{align}\label{Ricci3}
    \widehat{R}^a_{\ c}={R^a_{\ c}} \Rightarrow \widehat{R}=R\,.
\end{align}
Immediate consequence of equation (\ref{Ricci3}) is
\begin{align}
    \widehat{G}^a_{\ c}={G^a_{\ c}}\,.
\end{align}
In the extrinsically flat embedding, Gauss-Codazzi equations also simplify to
\begin{align}\label{Riemann3}
u^{[c}R_{a b m}^{\ \ \ \ d]}u^m = 2t_{[a}a_{b]}a^{[c}u^{d]} + 2u^{[c}(\nabla_{[a}a^{d]})t_{b]}\,.
\end{align}
So from equation (\ref{Riemann3}, \ref{Riemann}, \ref{S}, \ref{Weyl}) we get
\begin{align}
\widehat{R}_{a b}^{\ \ c d} = R_{a b}^{\ \ c d},~~\widehat{S}=S,~~\widehat{W}_{a b}^{\ \ c d}=W_{a b}^{\ \ c d}\,.
\end{align}
\\
\\
\noindent \underline{\textit{Corollary:}} \label{Claim-2} 
\textit{If $u^i=\xi^i / \sqrt{-g_{a b}\xi^a \xi^b}$ where $\xi^i$ is a hypersurface orthogonal  timelike Killing field, then the identities \ref{Identities} hold}. 
%
\\
\\
\textbf{proof:}
Let $N^2=-g_{a b}\xi^a \xi^b$. Then, the acceleration to the vector field $u^i$ can be written  $a_k=\nabla_k \ln N$, where we have used the fact $u^m\nabla_m N=0$ since $\xi^m$ is a Killing vector field. This immediately implies $K_{ab}=0$ \cite{Wald} \cite{static observer}, thereby proving the Corollary. 

\section{Examples}
We first discuss the example where the metric is time-independent. In this case, our results match with the usual Wick rotation. Then, we illustrate the time-dependent case, where there is no straightforward way to apply Wick rotation while still keeping the spacetime metric real.
\subsection{Accelerated Observers In Anti-de sitter Space}
One can consider a similar example for de-sitter space, though we considering the accelerated observers in Anti-de sitter space. The embedding equation of Anti-de Sitter space in a flat 5-dimensional space can be written as
\begin{align}
    − (z^0)^2 + (z^1)^2 + (z^2)^2 + (z^3)^2 − (z^4)^2 = −\ell^2 .
\end{align}
Global coordinates are provided by writing the general solution to the equation as,
\begin{align}
    z^0 = \ell \cosh{\rho}~\sin{\tau}, ~z^{\alpha}= \ell \omega^{\alpha}~\sinh{\rho}, ~z^4 = \ell \cosh{\rho}~\cos{\tau}\,,
\end{align}
where $\delta_{\alpha \beta}\omega^{\alpha}\omega^{\beta}=1$. Then one finds the metric
\begin{align}\label{Ads}
    ds^2 = \ell^2(− \cosh^2{\rho} d\tau^2 + d\rho^2 + \sinh^2 {\rho} d\Omega^2_2)\,,
\end{align}
with $0\leq \rho <\infty$ and $−\infty < \tau< +\infty$. Let us choose the space-time foliation by the observers whose  tangent vectors are always in the direction of the global timelike Killing vector,  $u^a=(\frac{1}{\ell \cosh {\rho}},0,0,0)$. These are clearly accelerated observers with $a_m=(0, \tanh{\rho}, 0, 0)$.
We write Ricci tensor, Ricci scalar, Einstein tensor respectively as follows,  using  $\nabla_m a^m=\frac{3}{\ell^2}, ~^{(3)}R=-\frac{6}{\ell^2}, K=0$.
\begin{align}
\widehat{R}^a_{\ c}=-\frac{3}{\ell^2}\delta^a_{\ c};~~\widehat{R}=-\frac{12}{\ell^2};~~ \widehat{G}^a_{\ c}=\frac{3}{\ell^2}\delta^a_{\ c}\,.
\end{align}
All are $\Theta$ independent as already discussed in the corollary(\ref{Claim-2}). We also see the Euclidean metric for (\ref{Ads}) is again maximally symmetric. 

\subsection{Accelerated Observers In Time-Dependent Spherically Symmetric Spacetime}
Any spherically symmetric metric can locally be expressed in the following form
\begin{align}\label{Kodama}
    ds^2=\gamma_{A B}(x^A)dx^A dx^B+\Tilde{r}^2(x^A)d\Omega^2,~~~~~ A,B \in \{0,1\}\,.
\end{align}
It is known that there exist  special fiducial observers called Kodama observers in any time-dependent spherically symmetric metric. Given the metric (\ref{Kodama}), it is possible to introduce the Kodama vector field $\textbf{k}$, those components are
\begin{align}\label{Kodama2}
    k^A(x)=\frac{1}{\sqrt{-\gamma}}\varepsilon ^{A B}\partial_B \Tilde{r},~~k^{\theta}=k^{\phi}=0.
\end{align}
From the above equation(\ref{Kodama2}) we conclude that the Kodama observers are characterized by the condition $\Tilde{r}=C(\Tilde{r_0})$, where $C$ is constant. And the remarkable corresponding conserved current is $J^a=G^a_{\ b}k^b$\cite{Kodama}.

Let us consider an example of metric (\ref{Kodama}) by considering the following metric of de Sitter space for a comoving  observer,
\begin{align}
    ds^2=-dt^2+e^{2 H t}(dr^2+r^2d\Omega_2^2)\,.
\end{align}
Consider the observers (Kodama Observers) stay at a fixed distance from its cosmological horizon move along the trajectory $r e^{H t}=C$(where $C$ is constant) with four-velocity in the direction of Kodama vector {$k^a=(-1, Hr, 0, 0)$},
\begin{align}
    u^a=\frac{k^a}{|\textbf{k}|} =\frac{k^a}{\sqrt{1-H^2C^2}}\,.
\end{align}
These observers foliation space(time) into orthogonal hypersurfaces with acceleration,
\begin{align}
    a_a=\left(\frac{H^3 C^2}{H^2C^2-1},\frac{H^2C^2r^{-1}}{H^2C^2-1},0,0\right)\,.
\end{align}
One can calculate the curvature tensor and  its  concomitants possessed by $\widehat g$ by using equations (\ref{Riemann}-\ref{Weyl}). We write the following
\begin{align}
    \widehat{R}=12H^2, ~ G_{a b}u^a u^b=3H^2\,.
\end{align}
There is a locally conserved current $J^a$  in terms of the Einstein tensor and the Kodama vector,
\begin{align}
    \widehat{J}^a=\widehat{G}^a_{\ b} k^b=(3H^2,-3H^3r,0,0)\,.
\end{align}
By using the relation (\ref{definition}), We write the metric as 
\begin{align}
    ds^2=-\left(1-\frac{F}{1-H^2C^2}\right)dt^2+\frac{2F~H^2C^2r^{-1}}{1-H^2C^2}dtdr+e^{2Ht}\left((1+\frac{F~H^2C^2}{1-H^2C^2})dr^2+r^2d\Omega^2_2\right)\,,
\end{align}
where $F={\Theta}/{(1+\Theta)}$. This gives the Euclidean metric with real entries for $F=2$ ($\Theta=-2$). Contrary to this, the usual Wick rotation gives the complex metric.

\section{Holonomy Along Closed Loop}

It has long been known that thermal effects associated with horizons can be understood in terms of holonomy about certain loops in the Euclidean spacetime, obtained by setting $t \to i t$, for a chosen time coordinate $t$. For example, for Rindler horizons in flat spacetime, $t$ is chosen to be the proper time of an accelerated observer, while in Schwarzschild, it is the time coordinate that appears in the standard form of the metric. More recently, in \cite{Samuel} it was shown that demanding the holonomy of null curves in the Euclidean spacetime to be trivial indeed gives the standard temperature associated with these spacetimes. Motivated by this, our aim is to study the holonomy of a special class of loops in spacetimes given by $\widehat g_{ab}$, particularly when the loop crosses the transition surface $\Sigma_0$ so that part of it lies in the Euclidean domain. Our setup a priori does not seem to bear any direct relation to the work in \cite{Samuel}, although it is in similar spirit. Moreover, there might be a curious connection that should be apparent from the final result and comments presented at the end of this section. 

Since accelerated observers play the central role as far as thermal effects are concerned, we need to consider $a^i \neq 0$. Consider a small rectangle with its sides given by $u^i$ and $S^m:={a^m}/{|a|}$. The area form associated with this loop is then given by $\Sigma^{m n}=u^{[m}S^{n]}$. 

\subsection{Loops in Euclidean regime}

To compute holonomy about such loops as mentioned above, it is easy to use the expression for the Christoffel connection $\widehat \Gamma^a_{\phantom{a} bc}$ given in Appendix \ref{app:data1}. We will discuss this in the next section, but before proceeding to that, in this section we analyze the standard expression for change of a vector, say $X^i$, about such a loop in terms of the curvature tensor. This should give a rough idea about the additional terms that might arise due to $\Theta$ and $\dot \Theta$ terms in the curvature tensor: 
$\widehat{\delta X^i} = {\widehat R}^i_{\ b c d} X^b \Sigma^{c d} \; \delta u \; \delta s$, where $\delta u$ and $\delta s$ are parameters along $u^i$ and $S^i$ respectively. From the previously established identities, it is easy to see that
\begin{align}\label{holonomy}
{\widehat R}^i_{\ b c d} X^b \Sigma^{c d}= R^i_{\ b c d} X^b \Sigma^{c d}&+\Theta \Biggl( -R_{a b c d}u^a u^i\Sigma ^{c d}+u^i\nabla _b |a|  -\frac{1}{1+\Theta}g^{a i}t_b \nabla_a |a|+ F t_b u^i  \nabla _{\vec{u}} |a| \nonumber\\
&~~~~~~~~+ S_b u^i |a|^2-\frac{1}{1+\Theta}t_b S^i |a|^2 \Biggl) X^b\nonumber\\
&+\frac{\dot{\Theta}}{2} \left( K_{b m}u^i S^m-\frac{1}{1+\Theta}t_b K^i_{\ m}S^m\right) X^b\,.
\end{align}
The above expression simplifies considerably in static spacetimes if one chooses $u^i$ in the direction of the timelike Killing vector. Using various standard identities (see, for example, \cite{static observer}), the above expression then reduces to 
\begin{align}\label{holonomy2}
    {\widehat R}^i_{\ b c d} X^b \Sigma^{c d} = R^i_{\ b c d} X^b \Sigma^{c d}-
    \underbrace{F S^i t_b X^b \left( |a|^2+S^m \nabla_m |a|\right)}_{\mathrm{additional~term}}\,,
    \hspace{0.9cm}
    \mathrm{(Static~Killing~Foliation)}
\end{align}
where $F={\Theta}/{(1+\Theta)}$ and $F=2$ in the Euclidean regime with $\Theta=-2$. 
The additional term above, which depends purely on acceleration, is worth exploring further in some physically relevant spacetimes. Let us consider
a static spherically symmetric spacetime, described by the standard line element
\begin{align}
    ds^2=-B(r)dt^2+\frac{1}{B(r)} dr^2+r^2 d\Omega^2\,,
    \label{eq:sph-symm}
\end{align}
where $B(r)$ is an arbitrary function such that $B'(r)$ 
vanish at infinity and $B(r)$ has a zero at some finite radius: $B(r_0)=0$.
In this case, the previous expression reduces to 
 \begin{align} \label{Holonomy3}
    {\widehat R}^i_{\ b c d} X^b \Sigma^{c d} = R^i_{\ b c d} X^b \Sigma^{c d}+(F/2) \left( S^i t_b X^b \right) \; {}^{(2)}R_{t,r}\,,
 \end{align}
where ${}^{(2)}R_{t,r} = - B''(r)$ is the curvature scalar of the two dimensional space $\theta, \phi=$ constant.
%
%

We will now highlight a possible connection of the additional term above with the relationship between Euclidean holonomy and temperature, in particular with the discussion in \cite{Samuel}. Let us choose our vector $X^i$ to be $u^i$ and imagine moving this vector about a loop in the Euclidean domain ($\Theta=-2, F=+2$) defined by a rectangular region in the $t-r$ plane bounded by $t=t_1, t=t_1+\beta, r=r_0, r=b$. Here, $\beta>0$ is a constant parameter and consider $b>r_0$ to be some large radius (below we assume $b \to \infty$). The area measure of such a loop is simply $dt dr$ (the $B(r)$ factor cancels out) and the integration of the last term in equation (\ref{Holonomy3}) gives
\begin{align}
   S_i \delta X^i = - \beta B'(r_0)\,.
\end{align}
This is an instructive result. For spacetimes of the above form ( \ref{eq:sph-symm}), the quantity $B'(r_0)=2 \kappa$ where $\kappa$ is the surface gravity of the horizon defined by $B(r_0)=0$. The RHS above is therefore of magnitude $2 \beta \kappa$. Now, the Hawking temperature associated with the horizon is $2 \pi T_{\rm H}$, therefore $\beta B'(r_0) = 2 \pi$, if one chooses $\beta = (2 T_{\rm H})^{-1}$.

The above analysis, though suggestive, leaves several unclear points, which we list below: 
\begin{enumerate}
    \item First, let us point out that while the last term equation (\ref{Holonomy3}) has been written in a nice geometric interpretation (with no approximations made), the connection we have highlighted with surface gravity and the range of time integration $\beta$ depends on choice of the vector and the loop. It is not clear how to interpret equation (\ref{Holonomy3}) for a generic case.
    \item The expression for change of vector in terms of Riemann tensor holds only for small loops, but we have here taken $b \to \infty$ so that the contribution from the $r=b$ vanishes. Essentially, what we have given is an interpretation for the contribution of this term due to the presence of the horizon at $r=r_0$.
    \item There is a factor $2$ mismatch between $\beta^{-1}$ and $T_{\rm H}$. This is puzzling, we do not know how this must be interpreted! The only place in the literature (as far as we are aware) where such a factor two discrepancy has been arrived at, by completely different set of arguments, is an old paper by Gerard 't Hooft \cite{thooft}. 
    \item The discussion above is tied to static horizons, but it is important to repeat it for stationary horizons to see how general is the result. This would require generalising the whole analysis to the case when $u^a$ is not hypersurface orthogonal. Some aspects of this are given in the Appendix \ref{NOH}, but the Riemann tensor will be more difficult to obtain.
\end{enumerate}


\subsection{Loops straddling the transition surface}

\begin{figure}[!h]
\begin{center}
\scalebox{0.45}{\includegraphics{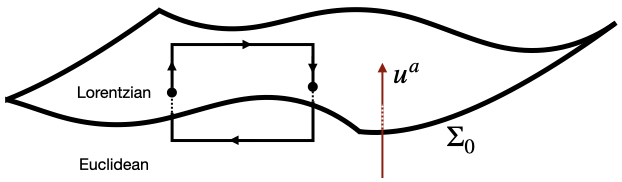}}
\end{center}
\vskip -5pt
\caption{Loops straddling the transition surface $\Sigma_0$.}
\label{fig:straddling-loops}
\end{figure}

As a more interesting case, we now comment on loops that straddle the transition surface $\Sigma_0$, so that part of these loops lie in the Euclidean regime; see Fig. \ref{fig:straddling-loops}. 

While using an analysis similar to the one in the preceding section, one must be careful since the metric $\widehat g$ is degenerate on $\Sigma_0$, therefore the area measure of the loop needs to be properly defined. However, a more immediate analysis can be presented in terms of the connection itself, which is given in Appendix \ref{app:data1}.

Let us choose our vector $X^i$ to be such that $X^i t_i=0$ everywhere in the region of interest and similarly, let $s^i$ be a properly normalized vector orthogonal to $u^i$. Imagine parallel transporting $X^i$ about the loop in Fig. \ref{fig:straddling-loops}, whose legs are defined by tangents $u^i$ and $s^i$. Then, we can estimate the change in the vector using the expression for the connection, which reads (see equation \ref{eq:connection} in (Appendix \ref{app:data1}):
\begin{eqnarray}
\widehat \Gamma^a_{\ b c} = \Gamma^a_{\ b c} + 
F \left[(1+\Theta) u^a K_{(b c)} - a^a t_b t_c \right]- (1/2) {\dot F} (1+\Theta) t_b t_c u^a
\end{eqnarray}
Above the surface, $\Theta=0=F$, while $\Theta=-2, F=+2$ below the surface. Therefore, the legs of the loop tangential to the surface will give different contributions to the change in vector and the additional contribution from the Euclidean domain is easily shown to yield
\begin{align}
  t_i \delta X^i = 2 \left( K_{ab} s^a s^b \right) \delta s\,,  
\end{align}
where $\delta s$ is the parameter along $s^i$. Although instructive, we are unable to say anything further about a generic interpretation of the above result. Moreover, we have assumed that the contribution of the legs normal to the surface can be made arbitrarily small (say, by letting $\delta u \to 0$). However, since the metric is becoming degenerate on $\Sigma_0$, how to handle the divergent $(1+\Theta)^{-1}$ terms is not very clear. At best, we can evaluate the above quantity in a simple spacetime such as the one in equation \ref{eq:sph-symm} with a suitable choice of $u^a$ and see if it yields anything sensible. For this purpose, we consider the region $r<r_0$ of this spacetime and describe this in new coordinate $\tilde t=r$, $\tilde r = t$, in which the metric becomes
\begin{align}
    ds^2=-\frac{1}{{\widetilde B}({\tilde t})} d {\tilde t}^2+{\widetilde B}({\tilde t})d{\tilde r}^2+{\tilde t}^2 d\Omega^2
    \label{eq:sph-symm-in}
\end{align}
where ${\widetilde B}({\tilde t}) = -B(\tilde t)$ and ${\tilde t}<r_0$. Thus, for Schwarzschild, we will have ${\widetilde B}({\tilde t}) = r_0/{\tilde t} - 1$. As before, we focus on the two dimensional plane with $\theta, \phi=$ constant. A trivial computation then gives 
\begin{eqnarray}
K_{{\tilde r} {\tilde r}} &=& \frac{1}{2} \sqrt{\widetilde B} \; \frac{\partial \widetilde B}{\partial \tilde t}~\,,
\;\;  \;\;\;  \;\;\;
\delta s = \sqrt{\widetilde B} \delta r\,,
\\
K_{ab} s^a s^b &=& \left( \frac{1}{2}\frac{\partial \widetilde B}{\partial \tilde t} \right) \delta r\,.
\end{eqnarray}
If we choose the transition surface as ${\tilde t}_0 = r_0 - \epsilon$ and evaluate everything at $\tilde t = r_0$, it is obvious that 
${\partial \widetilde B}/{\partial \tilde t} |_{{\tilde t}_0} = - 2 \kappa$ and the expression for change of vector now becomes $t_i \delta X^i = - 2 \kappa \beta$ with $\delta r = \beta$. This is the same as what we had obtained in previous section (the minus sign is easy to understand since here, the time coordinate $\tilde t$ {\it decreases} from $r_0$ to $0$ as we go into the Euclidean regime).

What we have sought to demonstrate in this section is a fascinating connection between holonomies about loops in space(time)s with distinct Euclidean and Lorentzian regimes. While the analysis is in the same spirit as the recent work in \cite{Samuel}, we must confess that a lot needs to be improved and several arguments need to be made rigorous, to get a complete picture based on our set-up. Nevertheless, the analysis above does show that one can extract quantities such as temperature very naturally by working within the completely covariant formulation given here, without having to consider complex values of time coordinate, very much in the spirit of the work in \cite{Samuel}.

\section{Euclidean actions and Entropy}

Having discussed the possible implication of our proposed covariant Wick rotation in the context of temperature associated with horizons in static spacetimes, we now investigate the issue of {\it entropy} in the same setting. This is expected to provide more non-trivial and interesting insights, since entropy associated with horizons depends on the explicit form of the Lagrangian of the theory under consideration, unlike temperature. 

Standard Euclidean techniques based on Wick rotation $t \to i t$ have been applied to obtain horizon entropy, essentially from the surface term in the gravitational action. We will briefly mention this towards the end of this section, but for now, we focus on another derivation which is motivated by the observation made by Visser in \cite{Visser}. The basis idea here is physically well motivated and yields an expression for entropy which matches with Wald entropy for a class of Lagrangians of the form $L(g_{ab}, R_{a b c d})$. We summarise the basic idea here and refer the reader to \cite{Visser} for further discussion. Let $L_{\rm E}$ be the Euclidean Lagrangian constructed from $L$ by Wick rotation, $t \to i t$, which is well defined for static spacetimes. Let $t_{ab}$ be the ``stress-energy" tensor defined by
\begin{eqnarray}
    I &=& \int L \sqrt{-g} d^4x \,,
    \nonumber \\
    \delta_{g} I &=& - \frac{1}{2} \int t_{ab} \delta g^{ab}\sqrt{-g} d^4x \,.
\end{eqnarray}
The object $t_{ab}$ is therefore the conventional metric stress-energy tensor if $L$ is the matter Lagrangian. However, one may define $t_{ab}$ similarly for gravitational Lagrangians as well, in which case we will obtain
\begin{eqnarray}
    t_{ab} = - 2 E_{ab} \;\;\; (\mathrm{gravitational~Lagrangian}) \,,
\end{eqnarray}
where $E_{ab}$ represents the gravitational equation of motion tensor; for example, for Einstein-Hilbert Lagrangian, $E_{ab}=(16 \pi G)^{-1} G_{ab}$. Given these definitions, the key observation made in \cite{Visser} is that the difference between $t_{ab} u^a u^b$ and $L_{\rm E}$ is a measure of entropy contributed by the fields with Lagrangian $L$. Although the discussion in \cite{Visser} separated out the Einstein-Hilbert part, as we will show below, this is not necessary. 

In this section, we will use the above set-up and check how it works when the Euclidean regime is defined by the $\Theta<-1$ domain of the metric $\widehat g$. We will see that, in general, the entropy obtained by using the above method with our covariant Wick rotation comes very close to the known results, matching them when $\Theta=-2$. However, in general, there are foliation dependent corrections that will appear in our case due to the presence of $K_{ab}$ in various expressions. Except for extrinsically flat foliations, such terms will not vanish. In particular, these terms will contribute for non-stationary horizons and hence may have physically relevant role to play in considerations such as generalised second law. 

To proceed with the calculation, we define, following \cite{Visser}, the so called ``anomalous" entropy as
\begin{eqnarray}\label{entropy}
    S_{\rm anomalous} = t_{ab} u^a u^b + L_{\rm E}\,.
\end{eqnarray}
The tag ``anomalous" was used in \cite{Visser} since, as mentioned above, that work focussed on {\it deviations} from the Bekenstein-Hawking entropy $S={A}/{4}$ in Einstein-Hilbert theory. We will keep the tag, but as we will see, there is no need to separate out the Einstein-Hilbert part. We will analyse the above expression for some well-known Lagrangians, thereby deduce their contribution to entropy.

\subsection{Scalar field theory}

We start with the simplest example of a scalar field theory in curved space time, with the Lagrangian and the stress-energy tensor given by standard expressions
 \begin{eqnarray}
     {L} &=& -\frac{1}{2}g^{a b}\nabla_a \phi \nabla_b \phi -V(\phi)\,,
      \\
     t_{ab} &=& \partial_a \phi \; \partial_b \phi - ({1}/{2}) g_{ab} \left(g^{ij} \partial_i \phi \partial_j \phi \right) - g_{ab} V(\phi)\,.
 \end{eqnarray}
 The above Lagrangian, for metric $\widehat g$, becomes
 \begin{eqnarray}
     {\widehat L} &=& -\frac{1}{2} {\widehat g}^{a b}\nabla_a \phi \nabla_b \phi -V(\phi)
     \nonumber \\
     &=& {L} + \frac{1}{2} \Theta \left( u^a \partial_a \phi \right)^2\,.
 \end{eqnarray}
 From the given expressions, it trivially follows that
 \begin{eqnarray}
 t_{ab} u^a u^b + {\widehat L}_{\Theta=-2} = 0\,.
 \end{eqnarray}
 Therefore, if we define $\mathcal{L}_{\rm E} = {\widehat L}_{\Theta=-2}$, we get
 \begin{eqnarray}
 S_{\rm anomalous}=0\,.
 \end{eqnarray}
 \\
 It is straightforward to establish the above analysis for more general scalar field Lagrangians, but it must be clear that, unless there are higher derivative terms and/or curvature couplings, the extrinsic curvature terms will not explicitly appear in the final result. 
 
 \subsection{Electromagnetic field theory}
 
For EM field, the Lagrangian and the stress-energy tensor are 
 \begin{eqnarray}
     {L} &=& - (1/4) g^{a c} g^{b d} \; F_{a b} F_{c d}\,,
     \\
     t_{ab} &=& - F_{am} F^{m}_{\phantom{m}b} + { L} g_{ab}\,.
 \end{eqnarray}
 For the metric $\widehat g$, the Lagrangian becomes
  \begin{eqnarray}
     {\widehat L} &=& - (1/4) {\widehat g}^{a c} {\widehat g}^{b d} \; F_{a b} F_{c d}
     \nonumber \\
     &=& {L} - \frac{1}{2} \Theta F_{am} F^{m}_{\phantom{m}b} u^a u^b\,.
 \end{eqnarray}
 Once again, if we define ${L}_{\rm E} = {\widehat L}_{\Theta=-2}$, we get $S_{\rm anomalous}=0$.

A more non-trivial example is given by the general vector field theory with action
 \begin{align}\label{vector field}
     I_{\rm vector}=\int (g^{a b}g_{m n}\nabla_a V^m \nabla_b V^n) \sqrt{-g} d^4x\,.
 \end{align}
 Although we do not analyse this action in detail, it is obvious that the Euclidean action will now have additional terms that might survive even when $K_{ab}=0$. For example, the above Lagrangian will have the additional terms of the form $V^j V^l$ contracted with
 \begin{align}\label{vector field 2}
     \left( g^{a b}g_{m n}+Fg^{a b}t_mt_n-\Theta g_{m n}u^a u^b-F\Theta u^a u^bt_m t_n\right) \left(C^m_{\ a l}C^n_{\ b j}+C^m_{\ a l}\Gamma^n_{\ b j}+\Gamma^m_{\ a l}C^n_{\ b j}+\Gamma^n_{\ b j}\Gamma^m_{\ a l}\right)\,,
 \end{align}
 with $C^m_{\ i j}$ given in the equation (\ref{eq:connection}). 
 

\subsection{Einstein-Hilbert}
 
We now apply the same method as above to gravitational Lagrangians, starting with the Einstein-Hilbert action $\mathcal{L}=(16 \pi G)^{-1} R$. As stated in the introductory paragraph of this section, in this case, $t_{ab}=-2 G_{ab}/(16 \pi G)$ and the Lagrangian $\widehat{\mathcal{L}} = {\widehat R}/(16 \pi G)$, where $G$ is the universal gravitational constant, we will use absolute units for the calculation purpose. Since we have already given the expression for $\widehat R$ in the equation (\ref{Ricci2}), upon using standard differential geometric identities, it is easy to prove that entropy density has additional foliation dependent terms:
\begin{eqnarray}
 S_{anomalous}&=&\mathcal{\rho}_L+{L}_E=-2G_{a b}u^a u^b+\widehat{R}\,, \nonumber \\ 
&=&2R_{a b}u^a u^b+2K_{m n}K^{m n}-4\nabla_m a^m-2K^2\,, 
\end{eqnarray}
where $\rho_L=t_{a b}u^a u^b$, $L_E$ is Euclidean Lagrangian constructed from Einstein-Hilbert Lagrangian by covariant Wick rotation. For static spacetime above expression reduce to 
\begin{align}\label{S1}
    S_{anomalous}=-2R_{a b}u^a u^b=-2\nabla_m a^m\,.
\end{align}
We get the associated entropy for static spacetime perceived by the accelerated congruence after  Integrating the above equation and using the fact, only spatial components of $a^i$ is nonzero and applying divergence theorem.
\begin{align}\label{S2}
    \mathcal{S}_{anomalous}=\frac{1}{2}\int S_{anomalous}\sqrt{-g} d^4x=\frac{A}{4}\,.
\end{align}
 The factor of 2 in denominator  appears due to the convention $t_{a b}=-2G_{a b}$. Our formalism gives usual entropy $=\frac{A}{4}$ law, only for the static spacetime where our covariant alternative to Wick rotation reduces to usual Wick rotation. 
  \subsection{Lanczos-Lovelock gravity}

One of the most direct higher curvature generalisations of the Einstein-Hilbert Lagrangian are the so called Lanczos-Lovelock (LL) Lagrangians, which become non-trivial in $D>4$ and share several features of the Einstein-Hilbert Lagrangian, in particular yielding equations of motion which are second order despite the appearance of higher curvature terms in the Lagrangian. These features arise from the very special structure of these Lagrangians, reviewed at length in \cite{LL}. We refer the reader to this review for derivation of various identities that we will use below.

In  $D-$dimensions, the LL Lagrangian is given by the sum:
\begin{eqnarray}
L &=& \sum_m c_m L_m\,.
\label{eqnx} \\
L_m^{(D)} &=& \frac{1}{16 \pi} \frac{1}{2^m} \delta^{a_1 b_1 \ldots a_m b_m}_{c_1 d_1 \ldots c_m d_m} R^{c_1 d_1}_{~ a_1 b_1} \cdots R^{c_m d_m}_{~ a_m b_m}\,,
\label{eq:ll-action}
\end{eqnarray}
where the tensor appearing in the right hand side  of the equation (\ref{eq:ll-action})  is the completely antisymmetric determinant tensor defined  as:
 \begin{eqnarray}
\delta^{i a_1 b_1 \ldots a_m b_m}_{j c_1 d_1 \ldots c_m d_m} = {\mathrm{det}} \left[ \begin{array}{c|ccc}
\delta^i_j & \delta^i_{c_1} & \cdots & \delta^i_{d_m} 
\\
\hline
\\
\delta^{a_1}_j &  & & 
\\
\vdots & & \delta^{a_1 b_1 \ldots a_m b_m}_{c_1 d_1 \ldots c_m d_m} &
\\
\delta^{b_m}_j & & &  \vphantom{\bigg{|}} \end{array} 
\right] \;
\end{eqnarray}
for $m \geq 0$.
 The lowest order terms, $m=0,1$ correspond to cosmological constant and the Einstein-Hilbert action respectively, as can be easily seen by expanding the alternating determinant. For $m=1$, $L_1=(16 \pi)^{-1} R$, the factor of $16 \pi$ in the definition of $L_m$ essentially changes the right hand side of equations of motion from the conventional $8 \pi T_{ab}$ to $(1/2) T_{ab}$. The equations of motion for a generic LL Lagrangian $L=\sum_m c_m L_m$ are given by the following two equivalent forms:
 \begin{eqnarray}
E^a_{b} &=& \sum_{m} {c_m E^{a}_{b(m)}}=\frac{1}{2} \; T^a_{b}\,,
\nonumber
\end{eqnarray}
where
\begin{eqnarray}\label{ELL}
E^i_{j(m)} &=& \frac{1}{16 \pi} \frac{m}{2^m} \delta^{a_1 b_1 \ldots a_m b_m}_{j_{~} d_1 \ldots c_m d_m} R^{i d_1}_{a_1 b_1} \cdots R^{c_m d_m}_{a_m b_m} 
- \frac{1}{2} \delta^i_{j} L_m
\nonumber \\
&=&  - \frac{1}{2} \frac{1}{16 \pi} \frac{1}{2^m} \delta^{i a_1 b_1 \ldots a_m b_m}_{j c_1 d_1 \ldots c_m d_m} R^{c_1 d_1}_{a_1 b_1} \cdots 
R^{c_m d_m}_{a_m b_m}\,.
\label{eom-lovelock}
\end{eqnarray}
We may now proceed with our computations in the following two steps:  
\begin{enumerate}
\item Compute the Euclidean LL Lagrangian: This is easily done by replacing $R^{a b}_{~ c d} \to {\widehat R}^{a b}_{~ c d}$ in the equation (\ref{eq:ll-action}) above.
\item Compute $E^i_j t_i u^j$.
\item Compute the difference between the above two quantities, hence compute $S_{\rm anomalous}$.
\end{enumerate}
Right at the outset, it is obvious that the resultant expression will differ from Wald entropy\cite{Wald2} due to the addition terms involving the extrinsic curvature tensor $K_{ab}$. We will discuss these terms momentarily. Before that, let us consider the trivial case of $K_{ab}=0$, applicable to, say, the case of static Killing horizons. Recall one of the equations of (\ref{Identities})
\begin{eqnarray}
{\widehat R}^{a b}_{~ c d} = R^{a b}_{~ c d} \,.
\end{eqnarray}
The Euclidean version of the Lagrangian is:
\begin{eqnarray}
{\widehat L}_m^{(D)} = {L}_m^{(D)}
\end{eqnarray}
One can obtain anomalous entropy for the case of $K_{a b}=0$,  $S_{\rm anomalous}={\widehat L}_m^{(D)} - 2 E_{00} $, where 
$E_{00}:=E^a_{~ b} t_a u^b\,. $
\begin{align}
    S_{\rm anomalous}=-2 {\mathcal R}_{00}
\end{align}
where $\mathcal{R}_{ab}$ is defined by $E^a_b = \mathcal{R}^a_b - (1/2) L \delta^a_b$ and ${\mathcal R}_{00} = \mathcal{R}^a_b t_a u^b$. It is obvious that $\mathcal{R}_{ab}$ is the analog of the Ricci tensor for LL models and reduces to it for $m=1$. The above expression is known to give correct entropy that matches with Wald entropy\cite{Wald2} for Lovelock gravity.

Let us now derive the general entropy relation for LL Lagrangian with non vanishing extrinsic curvature. Substituting the equation (\ref{Riemann}) into the equations (\ref{eq:ll-action}). The $m$th order LL Lagrangian for D- spacetime dimensions, becomes
\begin{align}\label{SLL}
{\widehat L}_m^{(D)} = {L}_m^{(D)}+L_K+L_{\partial K}
\end{align}

\begin{eqnarray}
    L_K &=&\alpha \delta^{a_1 b_1 \ldots a_m b_m}_{c_1 d_1 \ldots c_m d_m} \sum_{r=1}^{m}(-2\Theta)^r\binom{m}{r}K^{c_1}_{[a_1}K^{d_1}_{b_1]}\ldots K^{c_r}_{[a_r}K^{d_r}_{b_r]}R^{c_{r+1} d_{r+1}}_{a_{r+1}b_{r+1}}\ldots R^{c_m d_m}_{a_m b_m}\label{SLL2}\\
    L_{\partial K}&=& +4\Theta \alpha  m \delta^{a_1 b_1 \ldots a_m b_m}_{c_1 d_1 \ldots c_m d_m}  u^{[c_m}\nabla_{[a_m}K_{b_m]}^{d_m]} \huge{\textbf{(}} R^{c_1 d_1}_{~ a_1 b_1}  \cdots R^{c_{m-1} d_{m-1}}_{~ a_{m-1} b_{m-1}} \nonumber\\
    &+&\sum_{r=1}^{m-1(m>1)}(-2\Theta)^r\binom{m-1}{r}K^{c_1}_{[a_1}K^{d_1}_{b_1]}\ldots K^{c_r}_{[a_r}K^{d_r}_{b_r]}R^{c_{r+1} d_{r+1}}_{a_{r+1}b_{r+1}}\ldots R^{c_{m-1} d_{m-1}}_{a_{m-1} b_{m-1}}\huge{\textbf{)}}\label{SLL3}
\end{eqnarray}
Above equation gives Euclidean LL Lagrangian for $\Theta=-2$. We write the final expression for $m$th order anomalous entropy by using the equations (\ref{SLL}-\ref{SLL3}) and (\ref{ELL}) . 
\begin{align}
 &S_{\rm anomalous}=-2\mathcal{R}_{00} +S_K +S_{\partial K}
\end{align}
\begin{eqnarray}
S_K &=& \alpha \delta^{a_1 b_1 \ldots a_m b_m}_{c_1 d_1 \ldots c_m d_m} \sum_{r=1}^{m} 4^r\binom{m}{r}K^{c_1}_{[a_1}K^{d_1}_{b_1]}\ldots K^{c_r}_{[a_r}K^{d_r}_{b_r]}R^{c_{r+1} d_{r+1}}_{a_{r+1}b_{r+1}}\ldots R^{c_m d_m}_{a_m b_m}\\
S_{\partial K}&=&-8\alpha  m \delta^{a_1 b_1 \ldots a_m b_m}_{c_1 d_1 \ldots c_m d_m} u^{[c_m}\nabla_{[a_m}K_{b_m]}^{d_m]} \huge{\textbf{(}}R^{c_1 d_1}_{~ a_1 b_1}  \cdots R^{c_{m-1} d_{m-1}}_{~ a_{m-1} b_{m-1}} \nonumber\\
&&~+\sum_{r=1}^{m-1(m>1)}4^r\binom{m-1}{r}K^{c_1}_{[a_1}K^{d_1}_{b_1]}\ldots K^{c_r}_{[a_r}K^{d_r}_{b_r]}R^{c_{r+1} d_{r+1}}_{a_{r+1}b_{r+1}}\ldots R^{c_{m-1} d_{m-1}}_{a_{m-1} b_{m-1}}\huge{\textbf{)}}\,.
\end{eqnarray}
Where $\alpha=\frac{1}{16\pi}\frac{1}{2^m}$, $\binom{m}{r}=\frac{m!}{r!(m-r)!}$. This entropy relation is much more general in the sense, it contains the additional terms apart from the term that gives the Bekenstein-Hawking entropy for static Killing horizons in four dimensions. For future work, it would be interesting to compare the terms we obtain with similar terms arising in other approaches to computing entropy. The closest to ours seems to be the approach sketched in \cite{Solodukhin}. (Similar terms also appear, for instance, in the discussion of holographic entanglement entropy -- see \cite{Xi Dong, Myers, Camps}. However, there does not seem to be any obvious connection between our analysis and these approaches.) One distinctive feature of the additional terms in our expression for entropy is the presence of terms with derivatives of extrinsic curvature.

\section{Arbitrary Foliation (Non-Orthogonal Hypersurfaces)}
The discussion has done so far, except some comments is under the assumption of orthogonal foliation. Let us now consider arbitrary foliation. The notion of time for such types of foliation is subtle because time synchronization is much more difficult for these observers. Still, we can write mathematical formalism for a specified function $\Theta$. We write Ricci scalar associated by $\widehat g$ in terms of quantities associated  with $g$, under the assumption that the changes of function $\Theta$ are in the direction of the observer's tangent vector and their direction of acceleration (\ref{thetadot})
\begin{align} \label{Ricci Scalar NOH}
\widehat{R}=R &+ \Theta \left( K^2 +\nabla _m a^m + \nabla_{\vec{n}}K + F w_{a b}w^{a b}-R_{a c} u^a u^c \right) +\dot \Theta K \nonumber\\
&+\frac{f'}{2} \left( 2(1+\Theta) \nabla_m a^m + (1+\Theta)^2 a^2+ a^2  +a^2 f'- a^2 \Theta^2 \right) +a^m (1+\Theta) \nabla_m f'\,,
\end{align}
where $w_{m n}$ is rotation Tensor (anti-symmetric part of $K_{m n}$), $f$ is some smooth scalar and $f'=\frac{f}{(1+\Theta)^2}$(See \ref{NOH}).

\section{Implications and Discussion}

We have shown that the usual Wick rotation is mathematically inconsistent as it does not generate the Euclidean metric in general. We began by proposing a covariant approach (previously discussed in \cite{Dawood})  by considering a class of spacetime metrics \textbf{$\widehat{g}_{a b}$} derivable from a Lorentzian metric \textbf{$g_{a b}$}, timelike curves characterised by the tangent vector $\textbf{u}$ and a function $\Theta$ that interpolate between the Euclidean and Lorentzian regimes. The approach that we present here is mathematically well defined and physically acceptable.
Key highlights of the paper and its consequences are as follows.

\begin{enumerate}
    \item The most interesting consequence of the covariant version of Wick rotation appears when we apply it to compute the black hole entropy.
    \textit{Our formalism modifies the Bekenstein-Hawking entropy by foliation dependent terms even for a simple Einstein Hilbert action.} This result, to the best of our knowledge, has not been discussed before in the discussions using Euclidean methods. The result matches with the conventional expression for entropy $S={A}/{4}$ law in the static spacetimes.
    \item To understand how our approach work when used to study thermal effects associated with accelerated or black hole horizons, we also discussed the holonomy of some chosen vectors about the certain class of curves, including ones that straddle the transition surface separating Euclidean and Lorentzian domains. Interesting, the result comes quite close to the standard expression for surface gravity, except for a factor $2$ ambiguity.
    \item We  pointed out (from equations \ref{Ricci} and \ref{Ricci Scalar NOH}) that the boundary term to Einstein-Hilbert action is independent of  the acceleration of the observers. Moreover, the curvature tensors and its concomitants are equal in both regimes if one chooses a congruence  that foliates the spacetime into extrinsically flat hypersurfaces.
    \item All our results make it very clear, except when the spacetime foliation has vanishing extrinsic curvature, there is no valid reason to consider $R$ (or $-R$) to be the Euclidean Lagrangian. It is obvious that the additional terms will not only affect classical geometrical variables (as we have shown), but they may also affect quantum mechanically since the Euclidean action appears explicitly in the phase of the saddle point approximation to the ground state wave function of a system.
\end{enumerate}
 




\appendix
\section{Appendix}\label{appendix}
\subsection{Conventions}
The Latin indices a, b, . . . i, j . . . , etc., run over 0,..,n with
the 0-index denoting the time dimension and (1,..,n) denoting the standard space
dimensions. The Greek indices, $\alpha,\beta,...$, etc., will run over 1,..,n. Except when
indicated otherwise, the units are chosen with $c = 1$, $\hbar=1$. Lorentzian metric signature is $(-,+,+,+)$ and Euclidean metric signature is $(+,+,+,+)$. Curvature tensor is defined by the convention
\begin{align}
A^c_{\ ;ab}-A^c_{\ ;ba}=R^c_{\ dab}A^d\,.
\end{align}
And the convention for extrinsic curvature of hypersurface foliated by an arbitrary observer is
\begin{align}
K_{ab}=\nabla_a u_b + u_a a_b\,.
\end{align}

\subsection{Definitions and useful identities: \texorpdfstring{$u^a$}{\265} hypersurface orthogonal} \label{app:data1}

\begin{align}\label{definition}
\widehat g^{a b}=g^{a b}-\Theta u^a u^b
,~~\widehat g_{a b}=g_{a b}+F t_a t_b, ~~g_{ab}u^b=t_a,~~g_{a b}u^a u^b=-1\,.
\end{align}
\begin{align}
 F=\frac{\Theta}{1+\Theta},~~
\dot F=\frac{\dot \Theta}{(1+\Theta)^2}\,.
\end{align}
\begin{align}
\nabla_a F=-\dot Ft_a,~~\nabla_a \dot F=-\ddot F t_a-\dot Fa_a\,.
\end{align}
Where $u^m$ is observer's velocity vector, $a^m$ is  observer's acceleration vector i.e. $a^m=u^l\nabla_l u^m$ and $a_m=g_{m n}a^n$. The Christoffel connection is given by
\begin{eqnarray}
\widehat \Gamma^a_{\ b c} &=& \Gamma^a_{\ b c}+C^a_{\ b c}\,,
\nonumber \\
C^a_{\ b c} &=& F \left[(1+\Theta) u^a K_{(b c)} - a^a t_b t_c \right]- (1/2) {\dot F} (1+\Theta) t_b t_c u^a\,.
\label{eq:connection}
\end{eqnarray}
The Riemann tensor associated with the above connection is given in the main text. Here, we quote a few alternate forms for the same which are helpful in simplifying certain expressions.
\begin{align}
    \widehat{R}^i_{\ b c d}=R^i_{\ b c d}&+2\Theta\Biggl( K^i_{\ [c}K_{d]b} - 2t_{[c}a_{d]}u^i a_b+\frac{2}{1+\Theta}(t_b t_{[c}\nabla^i a_{d]}-u^it_{[c}\nabla_b a_{d]}-a^i t_b a_{[c}t_{d]})\nonumber\\
    &~~~~~~~~~~~~~~~~~~~~~~~~~~~~~~~~~~~~~~~~~-F u^i t_b t_{[c}\nabla_{\vec{u}}a_{d]}-F u^i t_{[c}\nabla_b a_{d]} \Biggl)\nonumber\\
    &-\frac{\dot \Theta}{1+\Theta}\left(u^i t_{[c}K_{d] b}-t_{[c}K_{d]}^{\ i}t_b+\Theta t_{[c}K_{d] b}u^i\right)\,.
\end{align}
Above form is useful in the calculation for equations (\ref{holonomy}) and (\ref{tidal tensor}). Few steps using Gauss-Codazzi equations allows us to write the curvature tensors completely in terms of the extrinsic curvature and its derivative.
\begin{eqnarray}
    \widehat{R}_{a b}^{\ \ c d}&=&R_{a b}^{\ \ c d}+2\Theta\left(2u^{[c}\nabla_{[a}K_{b]}^{\ \ d]}-K_{[a}^{\ d}K_{b]}^{\ c}\right)+2\dot \Theta u^{\textbf{[}c}K_{[a}^{\ d\textbf{]}}t_{b]}\,,\\
    \widehat{R}_a^{\ c}&=&R_a^{\ c}+\Theta\left(u^c\nabla_a K-u^c\nabla_b K_a^{\ b}-u^b\nabla_a K_b^{\ c}+u^b\nabla_b K_a^{\ c}-K_a^{\ b}K_b^{\ c}+K K_a ^{\ c}\right)\nonumber\\
    &&\hspace{.9cm}+\frac{\dot\Theta}{2}(K_a^{\ c}-t_a u^c K)\,,\\
    \widehat{R}&=& R+\Theta(2\nabla_{\vec{u}}K +K_{m n}K^{m n}+K^2)+\dot\Theta K\,.
\end{eqnarray}
\subsection{Definitions and useful identities: \texorpdfstring{$u^a$}{\265} not hypersurface orthogonal}\label{NOH}
In this case, the basis definition of $F$ and $\dot F$ remain the same as above, but the key differences arise in the gradient of various functions
\begin{align}\label{thetadot}
&\nabla_a\Theta=-\dot\Theta t_a+f a_a,~~ \nabla_a F=-\dot F t_a + f' a_a \,,
\end{align}
\begin{align}
\nabla _m \dot F= f' A_m +f' a_b \nabla_m u^b-(\dot F-\dot f')a_m -\ddot F t_m\,.   
\end{align}
Where $f$ is some smooth scalar and $f'$ and $A_m$ are given as follows
\begin{align}
f'=\frac{f}{(1+\Theta)^2},~~A_m =\nabla_{\vec{u}} a_m \,.
\end{align}
And the difference in Christoffel connection $\Acute{C}^a_{\ b c }$ for Nonorthogonal case is given in terms of $C^a_{\ b c}$ of Orthogonal foliation,
\begin{align}
\Acute{C}^a_{\ b c}=C^a_{\ b c}-F K^{a }_ { \ (b} t_{c)}+F K_{(b}^{\ a}      t_{c)}+f'(a_{(b} t_{c)} u^a (1+\Theta)-a^a t_b t_c)\,.
\end{align}
Unlike the orthogonal case, $K_{m n}$ is not symmetric here i.e. $K_{m n}=K_{(m n)}+w_{m n}$


\end{document}